\documentclass{ptapap}

\author{Thomas Rivinius}[triviniu@eso.org,ESOC]
\author{Dietrich Baade}[dbaade@eso.org,ESOG]
\author{Alex C.~Carciofi}[carciofi@usp.br,IAG]

\affil[ESOC]{ESO --- European Organisation for Astronomical Research in the
  Southern Hemisphere, Casilla 19001, Santiago 19, Chile}
\affil[ESOG]{ESO --- European Organisation for Astronomical Research in the
  Southern Hemisphere, Karl-Schwarzschild-Str.~2, 85748 Garching, Germany}
\affil[IAG]{Instituto de Astronomia, Geof\'isica e Ci\^encias Atmosf\'ericas,
  Universidade de S\~ao Paulo, 05508-900, S\~ao Paulo, SP, Brazil}

\title{Be Stars Seen by Space Photometry}

\begin{document}

\maketitle

\begin{abstract}
Classical Be stars are introduced as object class and their particular
potential for space based photometry is highlighted. A brief summary of the
various types of variability observed in Be stars makes clear that an
interpretation of every single frequency as a pulsation mode falls short,
instead there are as well purely circumstellar variations and those that
originate in the immediate stellar to circumstellar interaction region. In
particular the latter offer great potential, as they are linked to one of
the few remaining great riddles of Be stars, namely how they feed their disks.
\end{abstract}

\section{Introduction}
The class of Be stars is simply defined as those non-supergiant B stars that
at least once have shown Balmer line emission \citep[see][for a
  review]{2013A&ARv..21...69R}. While this definition is very valuable for
bulk classification of stars for which no high quality data is available,
e.g., in the Magellanic Clouds, it is also very broad: Any circumstellar gas
close to a B star and above some threshold density will produce line
emission. In an attempt at a taxonomy that reflects our knowledge on how the
gas was put and is kept close to the star (e.g., magnetically confined, an
accretion disk, or a decretion disk), the class was subdivided. This
contribution focuses on the so-called classical Be stars.

Classical Be stars are rapidly rotating B stars, surrounded by a gaseous disk
that is formed by the star itself through mass ejection. As soon as there is a
disk, its fate is governed by viscosity, possibly with some contribution from
radiation through disk ablation. As a class, they are known to pulsate in
non-radial modes, but not to harbor large scale magnetic fields. In general,
Be stars and their disks could be considered as fairly well understood, was it
not for the one central question that is still open: How are these disks
formed and fed? Space photometry, combining long time bases, short cadence,
and high precision offers a new, unique, and highly promising approach to
answer this.

\subsection{Be stars and space photometry} 

Several space photometry missions, mostly designed for asteroseismology and
planet hunting, have been launched over the past decade. All of them have also
observed Be stars, partly in dedicated runs, as the MOST mission, but mostly
as by-catch to their regular observing strategies (CoRoT, Kepler, and also the
solar mission SMEI). The BRITE-Constellation observing strategy, while not
really focused on Be stars, is located somewhere between both approaches, as
it gives a certain favor to Be stars, owing to their brightness and relative
abundance among bright stars. 

To date MOST has observed five Be stars for several weeks each
\citep{2005ApJ...623L.145W,2005ApJ...635L..77W,2007ApJ...654..544S,2008ApJ...685..489C},
Kepler three for four years \citep{2016A&A...593A.106R}, CoRoT nearly 40 for
between a few weeks and half a year \citep[][and Rivinius et al., in
  prep]{2006A&A...451.1053F, 2007A&A...476..927G, 2009A&A...506...95H,
  2009A&A...506..125D, 2012A&A...546A..47N, 2009A&A...506..133G,
  2009A&A...506..143N, 2013A&A...551A.130S,2012PhDT.Semaan,
  2010A&A...522A..43E}, and SMEI almost 130 over nine years \citep[][and
  Rivinius et al., in prep]{2011MNRAS.411..162G,2014MNRAS.440.1674H}. BRITE is
still observing a growing list of Be stars \citep[published so
  far:][]{2016A&A...588A..56B}, and so is K2 (no published studies yet).

Each of these missions has its advantages and drawbacks when it comes to
understanding Be stars, and in the following it will be shown how the
photometric observables are formed in Be stars, what is hoped to be learned
from them, in particular concerning the unsolved problem of disk feeding, and
how the BRITE-Constellation finds its place in this roster of space missions
\citep[see][and Baade et al., this proceedings]{2016arXiv161002200B}

\section{Formation of photometric observables}

\begin{figure}
 \parbox{\textwidth}{
\parbox{0.03\textwidth}{~}\parbox{0.3\textwidth}{\centerline{31\,Peg}}%
\parbox{0.03\textwidth}{~}\parbox{0.3\textwidth}{\centerline{$\mu$\,Cen}}%
\parbox{0.03\textwidth}{~}\parbox{0.3\textwidth}{\centerline{$\omega$\,CMa}}%
 
\parbox{0.33\textwidth}{\includegraphics[width=0.33\textwidth,clip]{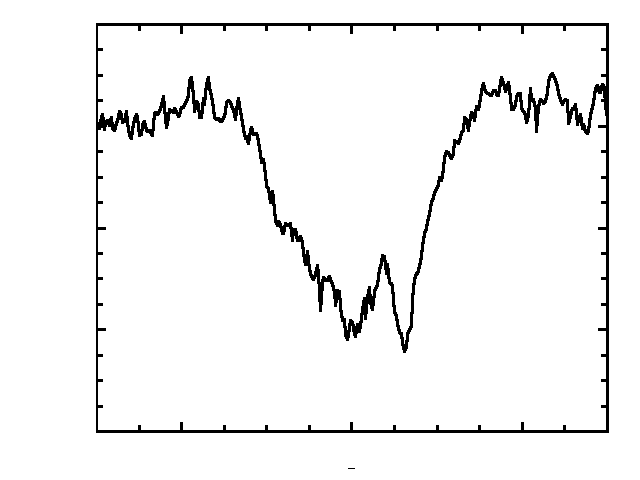}}%
\parbox{0.33\textwidth}{\includegraphics[width=0.33\textwidth,clip]{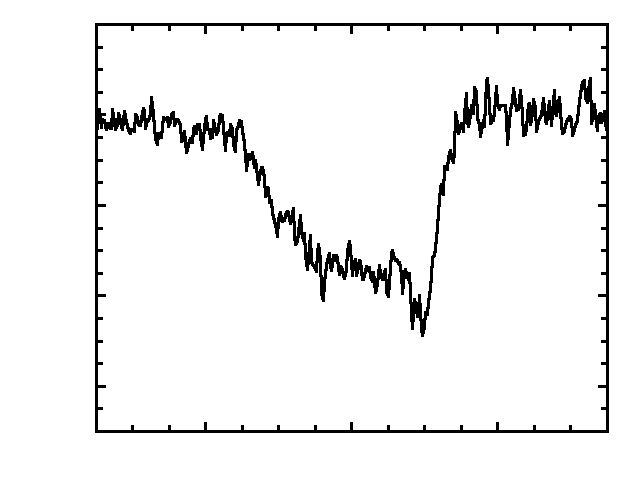}}%
\parbox{0.33\textwidth}{\includegraphics[width=0.33\textwidth,clip]{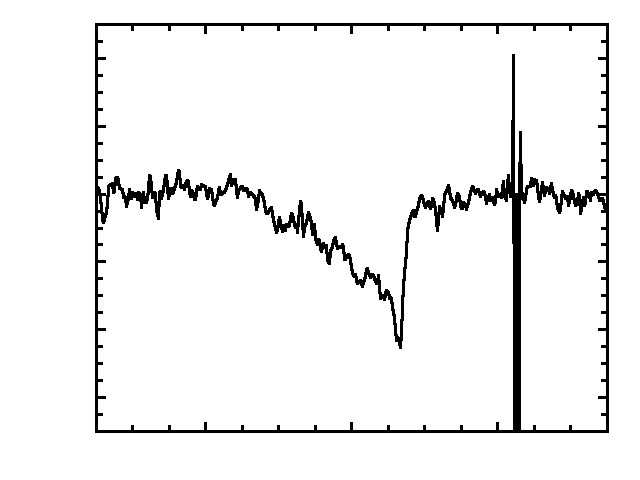}}%
}
   \caption{Example spectral line profiles of three low inclination Be stars at the
     pulsational phase in which the ``spike'' is most pronounced.}
   \label{fig:Be_spikes}
   \end{figure}

\subsection{Stellar pulsation}\label{sec:puls}
Classical Be stars, with very few possible exceptions, are pulsating stars,
which made them attractive targets for asteroseismology missions in the first
place. All published results from space missions (see above) have found
non-radial pulsation modes, with typical frequencies between 0.5 and
2\,c/d. Some earlier type stars show additional high frequencies. This picture
is commonly interpreted such that Be stars are mostly $g$-mode pulsators, with
some also showing $p$-modes. The associated light variations are due to
adiabatic changes in the local stellar surface emissivity, as the pulsational
waves pass over the stellar disk. This picture had been anticipated
spectroscopically, at least for the early type Be stars
\citep{2003A&A...411..229R}.

Mode identification from photometry alone suffers from the rapid rotation of
Be stars, well above 50\% of the critical fraction, which is beyond the
current limits of mode excitation theory. That said, most works based on
asteroseismology agree that Be stars are seen to pulsate in prograde, low
degree ($\ell\leq 4$) $g$-modes.

The spectroscopic picture is different. In particular Be stars seen at low
inclination, i.e., near pole on, show distinct pulsational features that (see
Fig.~\ref{fig:Be_spikes}), as long as one assumes traditional pulsation modes,
and not Rossby waves or others, can only be explained with retrograde
modes. These spikes arise due to a combination of a relatively low co-rotating
pulsational frequency, which causes the surface parallel velocity field to be
much higher than the surface normal one, with the the low inclination. For
this to happen, the mode has to be retrograde; prograde modes always have
co-rotating frequencies {\em higher} than the rotational one.  While
rotational velocity is projected with a low $\sin i$, the $\theta$ velocity
field is projected with a factor of, or at least close to, unity, and so the
two projected fields become comparable in magnitude. Adding up the velocity
fields, a large fraction of the stellar surface is projected into a narrow
velocity range, causing the spike \citep[see][for
  details]{2001A&A...369.1058R,2002Msngr.108...20R,2003A&A...411..181M}. Spectroscopically,
almost all Be stars seem to pulsate in a dominant $g$-mode, which turns out to
be $\ell=m=+2$, i.e., retrograde.

\subsection{Photometric contribution by the disk}\label{sec:disk}

\begin{figure}

      \includegraphics[width=.5\textwidth,viewport=80 120 432
        392,clip]{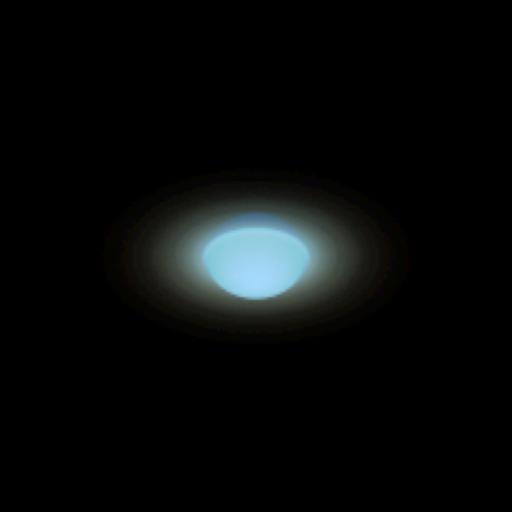}%
      \includegraphics[width=.5\textwidth,viewport=80 120 432
        392,clip]{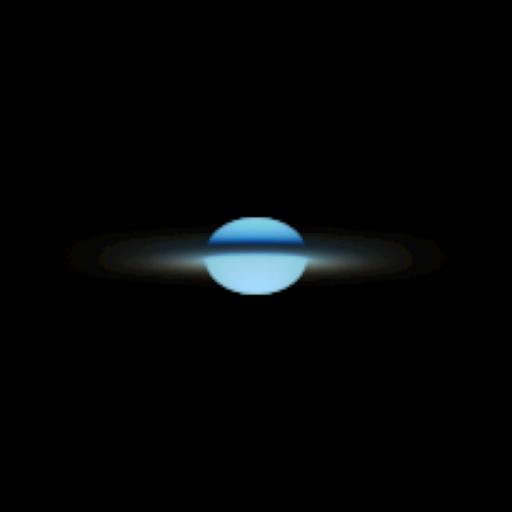}%

   \caption{Model of a Be star with a small disk, at two different
     inclinations, 56 and 84 degrees, as it would appear to the human eye
     (parameters like those of the B4\,IVe star Achernar, except the
     inclination). The disk structure was computed according to the viscous
     decretion disk model, the radiative transfer with the Monte Carlo code
     HDUST (Faes et al., in prep.)  }
   \label{fig:Be_hdr}
   \end{figure}

The circumstellar disk reprocesses the stellar light and in the process alters
the spectral energy distribution (SED). Depending on the inclination angle and
the wavelength the disk may increase the total flux for pole-on systems, and
decrease it for equator-on systems (see Fig.~\ref{fig:Be_hdr}). In the visual
domain, the inclination angle at which one regime changes into the other, and
hence the total flux remains unaffected by a disk, is at about 70$^\circ$
\citep[see][for general modeling of the disk
  photometry]{2012ApJ...756..156H}. The amount and precise properties of this
re-processing of light depend, apart from inclination, on the density and
density profile of the disk and the stellar SED irradiating the disk. The only
variable parameters, in a given system, usually are the density-related
quantities. The so-called light-house effect, in which it was speculated that
the stellar pulsational lightcurve is measurably re-processed in the disk was
not observed so far. Also, while Be stars with precessing disks exist, they
are very rare. Among Be stars brighter than $V=6.5$\,mag only three are known
($\gamma$\,Cas, 59\,Cyg, and 28\,Tau; all binaries), and only in one of them,
28\,Tau, the precession seems permanently present \citep{1998A&A...330..243H}.

When a Be star is in a diskless state and starts growing a disk, the disk
forms inside out, i.e., very close to the star the density rises quickly, but
further out only with a delay that depends on the viscosity and the
distance. After some time the disk reaches a steady state density profile,
that is best approximated by a radial power law $\rho\propto r^n$ with
$n=-3.5$ to $-3.0$. During the built-up phase the profile is steeper, and when
the disk decays it decays faster at smaller radii, leading to a shallower
density profile \citep[see Fig.~1 of][]{2012ApJ...756..156H}. 

For a given photometric band, the disk acts like a pseudo-photosphere of
radius $\bar{R}(\lambda)$, where most of the emitted flux comes from, surrounded by
optically thin material. The size of the pseudo-photosphere grows with
wavelength typically as $\lambda^{0.4}$ \citep{2015MNRAS.454.2107V}.  As a
result,
the color behaviors of growing vs.\ decaying disks are quite different, at
least for the stars with non-equatorial inclinations at which the disk absorbs
more light than it emits towards the observer \citep[see Fig.~2 and upper row
  of Fig.~12 of][]{2013A&ARv..21...69R}.

It is important to note that such secular changes of disk growth and decay,
even in the very close vicinity of the star, take at least several days, but
also may be much slower and be of the order of decades. This 
behavior, together with a typical lack of periodicity and phase coherence,
sets the photometric changes due to the disk density well apart from those
originating in the stellar photosphere, in other words they are,
photometrically, easy to distinguish.

While all Be stars, by definition, have a photon reprocessing disk in their
active state, and almost all Be stars show pulsational variability, many of
them also show additional variability that is not straightforwardly explained
in the above framework. It is tempting to link this, as of yet unexplained,
photometric behavior to the as of yet unexplained mass ejection process
that forms the disk.

In the following section, variability properties of Be stars are shown for
several typical cases.

\section{Types of Be star photometric variability}

\begin{figure}

      \includegraphics[width=.5\textwidth,clip]{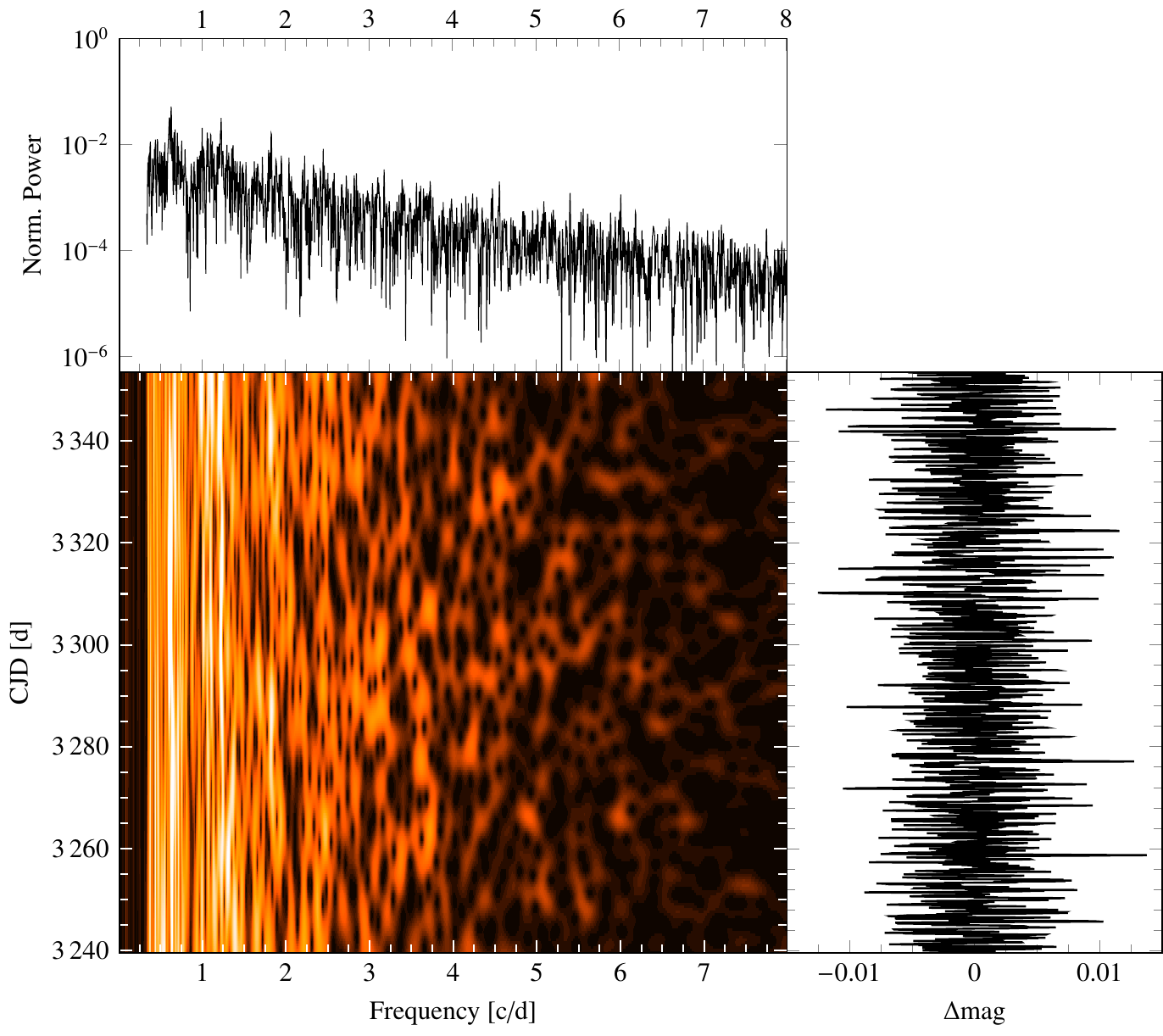}%
      \includegraphics[width=.5\textwidth,clip]{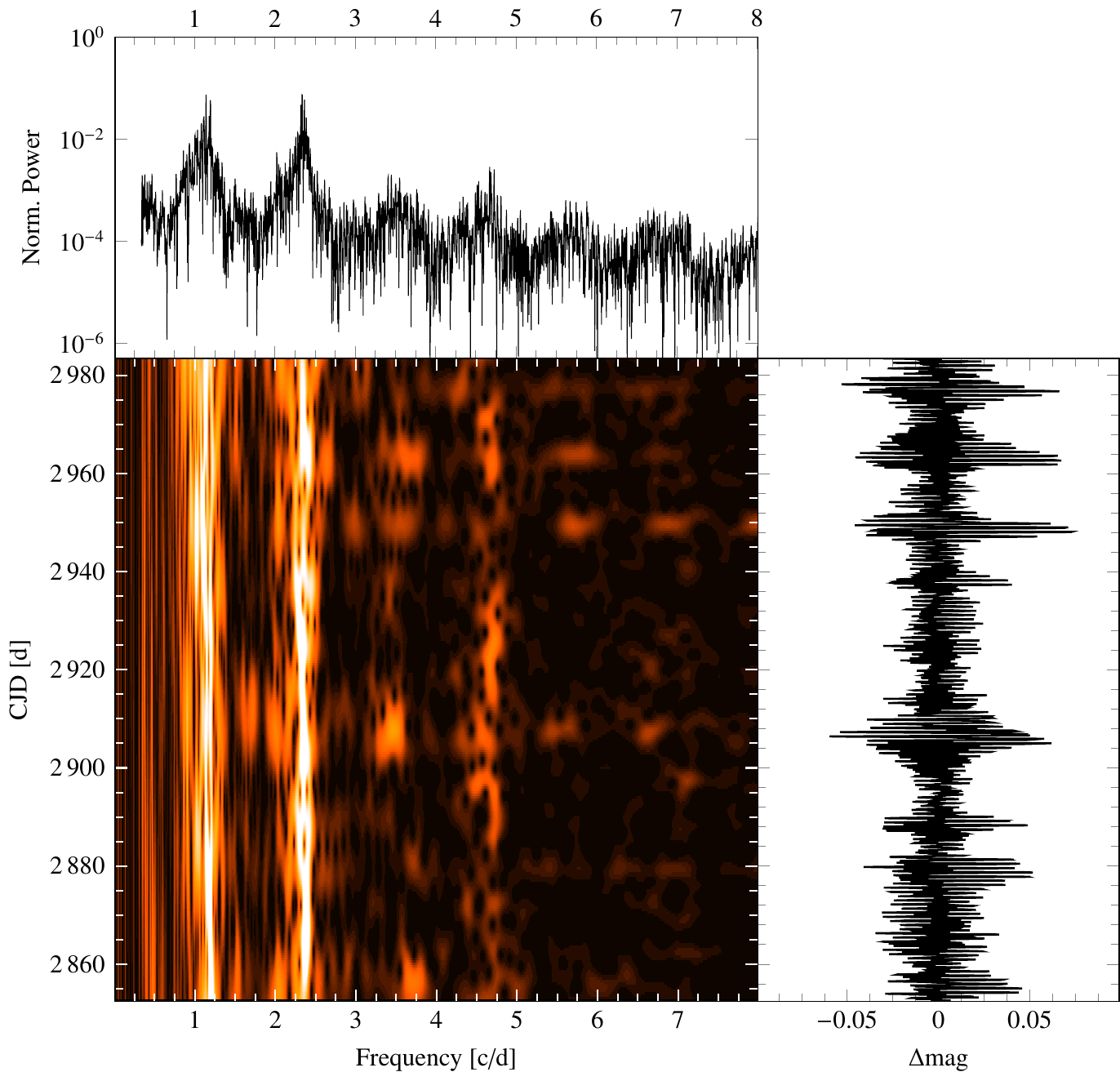}%

   \caption{Wavelet analysis (middle panels), Lomb-Scargle periodogram (top
     panels), and de-trended light curves (right panels) for HD\,51452 (left)
     and CoRoT 102719279 (right). From Rivinius et al., in prep., but see also
     \citet{2016A&A...593A.106R} for an introduction of the analysis methods
     used here.}
   \label{fig:Be_early}
   \end{figure}

\noindent {\bf Very early type Be stars:} HD\,51452 is a B0.5\,IVe star with
an emission appearance reminiscent of $\zeta$\,Oph, i.e. weak emission peaks
on either side of the rotationally broadened absorption profile.  The CoRoT
data for this B0\,IVe star were analyzed by \citet{2012A&A...546A..47N}.
Although the overall peak-to-peak variability with $\sim$20\,mmag is quite
strong, neither the power spectrum nor the wavelet analysis show clear
frequency peaks above a continuum of variability (see Fig.~\ref{fig:Be_early},
left). \citet{2012A&A...546A..47N} interpreted this as a sign of
stochastically excited pulsation in a multitude of modes, that arise and decay
on short time-scales. This behavior is not very common among Be stars, yet it
is not unique: $\gamma$\,Cas may show a similar behavior, as unpublished SMEI
data suggests.

\noindent {\bf Early type Be stars:} CoRoT 102719279 is a B1\,V shell star
with a strong and asymmetric H$\alpha$ emission. The star shows a large
number of small outbursts in the wavelet analysis, which all fainten the star
to a varying degree. The stellar brightness recovers quickly within a few days
or maximally two or three weeks after outburst.

The star is close to a textbook example of a Be (shell) star showing outbursts
(see Fig.~\ref{fig:Be_early}, right). The power spectrum out of outburst is
dominated by long-term persistent single, rather well defined
frequencies. During an outburst this changes dramatically, and in addition
strong and broad ``bumps'' of power around a base frequency and around a large
number of harmonics of that base frequency appear. It is worth reminding that
in shell stars additional disk material absorbs and scatters stellar radiation
out of the line of sight, so the star is comparable to $\eta$\,Cen, BRITE data
of which were analyzed by \citet{2016A&A...588A..56B}, and a bit like
StH$\alpha$166 \citep{2016A&A...593A.106R}, although there the individual
outbursts are not as well separated as here.

\begin{figure}

      \includegraphics[width=.5\textwidth,clip]{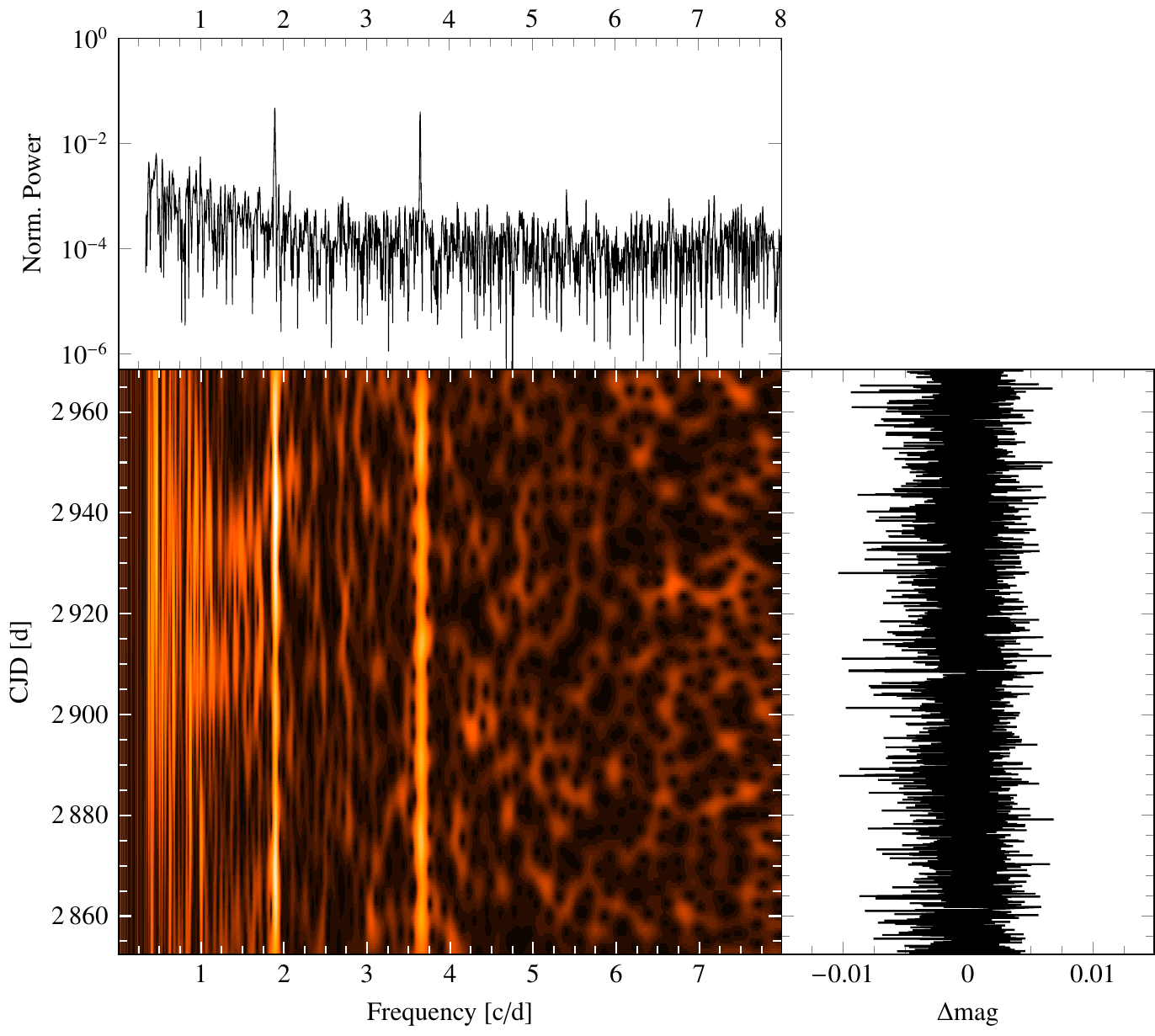}%
      \includegraphics[width=.5\textwidth,clip]{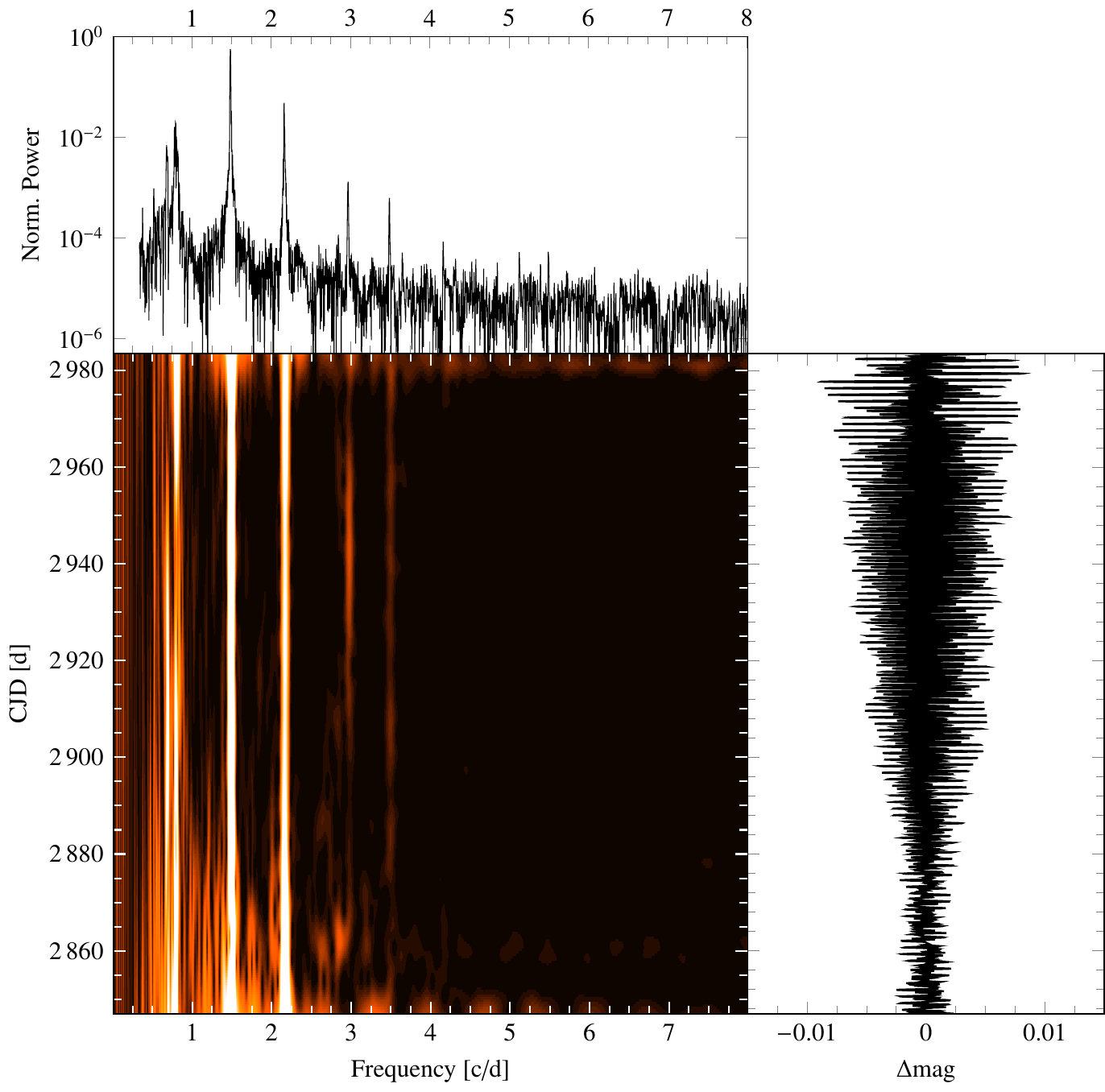}%

   \caption{As Fig.~\ref{fig:Be_early}, but for CoRoT 102721109 (left) and
     HD\,50209 (right). }
   \label{fig:Be_late}
   \end{figure}

\noindent {\bf Mid type Be stars:} CoRoT 102721109 is a mid to early type Be
star, but unfortunately a more precise spectral classification than being
hotter than 15\,kK is not available \citep{2013A&A...551A.130S}.  It shows no
sign of outbursts during the observations, and there are only two frequencies,
seemingly of stellar origin (see Fig.~\ref{fig:Be_late}, left).  Although the
higher frequency is not far from the harmonic of the lower one, it differs from
the precise harmonic value by about 4\%, which is a very significant
difference, given the time base and frequency resolution of the data.

\noindent {\bf Late type Be stars:} CoRoT data of the late type Be star
HD\,50209 (B8\,IVe, see Fig.~\ref{fig:Be_late}, right) was analyzed by
\citet{2009A&A...506..125D}.  The power spectrum is quite typical for its
position in the Hertzsprung Russell diagram: It shows no signs of outbursts
during the observing campaign. While many frequencies are apparent, most are
commensurate in that they could be explained as possible harmonics of a
frequency at about 0.74\,c/d. However, there is no frequency group at this
base value. Instead, the two lowest frequencies bracket that hjypothetical
base value on either side. Such a pattern, namely that there seems to be a
large number of harmonics, but are not confirmed in a more detailed analysis
is common in mid and late type Be stars.

\section{Discussion}

The four categories of variability described above, namely stochastic,
bursting, cleanly pulsating, and almost harmonics, are labeled with the
spectral subtype range in which they are common, but the borders between them
are not sharp.  The bursting behavior described under ``early type'' is the
most interesting for understanding the causes and properties of the disk
feeding, as outbursts, i.e., discrete feeding events, are clearly identifiable
and can be investigated in detail.

The most interesting result to date is the identification of difference
frequencies, see \citet{2016A&A...588A..56B} and Baade et al.\ in this volume,
but as well in many CoRoT targets (Rivinius et al., in prep.). The light
modulation with the difference frequencies is unlikely of stellar origin,
because the periods are too long for pulsational processes (see
Sect.~\ref{sec:puls}). More plausible is a swelling up and down of the density
of the innermost parts of the disk, affecting the light reprocessing (see
Sect.~\ref{sec:disk}). The parent frequencies of the difference frequencies
{\em are} pulsational, though. This means difference frequencies are a direct
sign that interaction between two pulsation modes does modulate the density
above the photosphere, and hence the disk feeding. A similar relation, between
pulsation and outburst has previously been identified in $\mu$\,Cen
\citep{1998cvsw.conf..207R,1998A&A...336..177R}. One might reconcile the two,
outbursts vs.\ swelling up and down, in a picture in which the same mechanism
acts in both types, but in $\mu$\,Cen is less efficient: There it would cross
a critical threshold to eject matter only at times of strongest
interaction. In other stars with difference frequencies, the ejection
mechanisms would always be active, but depending on beating phase at times
with lower or higher efficiency, and thus modulate the density (and light
reprocessing) of the innermost part of the disk in a more steady way.

In any case, past and ongoing asteroseismic space missions have enabled the Be
star community to come forth with well advanced hypotheses concerning the mass
ejection, that make predictions about when outbursts should occur and how they
should look like, which are well testable by ground-based follow-up
observations. This might prove the final break-through to solve the
long-standing question of disk feeding in Be stars.

\section{The role of BRITE-Constellation}
What place does BRITE-Constellation have in investigating Be stars? All the
missions acquiring that type of data have unique properties. Kepler, K2, MOST,
and CoRoT deliver data quality in which every individual cycle can be traced,
and with which BRITE cannot compete.  However, MOST and K2 campaigns are much
shorter than BRITE ones, Kepler targets are virtually unknown stars otherwise,
and CoRoT targets are only marginally better investigated, while BRITE
observes stars for which usually a large body of supporting studies already
exist, and which are even in the reach of the amateur spectroscopic
community. SMEI has a hardly challengable time base of nine years and
concentrates on a similar brightness range as BRITE-Constellation, but the
data quality is so limited that the only reason pulsation frequencies are
found at all is because cycles repeat several thousand times during the time
base. No individual outbursts can be identified in SMEI data.

While BRITE-constellation does not outshine these spacecraft in any particular
single mission parameter, in the combination of the observational properties
it is the best possible compromise: It is more precise than SMEI, provides
longer time bases than MOST and K2, and its targets are much better known Be
stars than those observed by Kepler or CoRoT will ever be. The latter is a key
point.  Be stars are variable on a large range of time scales. At the very
least the build-up or decay state of the disk must be known for a
meaningful interpretation of the observed variability, and such and other
supporting observations are much easier to acquire for the brighter Be stars
observed by the BRITE-Constellation mission.


\bibliographystyle{ptapap}
\bibliography{Rivinius}

\end{document}